\documentclass[11pt,letterpaper]{article}

\title{
\bf Geometrodynamics and Lorentz symmetry}
\author{
{\bf Derek K.\ \!Wise} \\[.5em]
{\sl \small Institute for Quantum Gravity} \\[-.3em]
{\sl \small Universit\"at Erlangen--N\"urnberg} \\[-.3em]
{\sl \small Staudtstr.\ \!7/B2,\! 91058 Erlangen,\! Germany} \\
\small \texttt{derek.wise@gravity.fau.de} 
}

\usepackage{amssymb,amsmath,amsthm}

	\makeatletter 
	\def\th@plain{%
	  \thm@notefont{}
	  \itshape 
	}
	\def\th@definition{%
	  \thm@notefont{}
	  \normalfont 
	}
\makeatother

\usepackage[
    colorlinks,%
    linkcolor=blue,citecolor=red,urlcolor=blue,
]{hyperref}
\usepackage[all,knot]{xy}
\usepackage{tikz}
\usetikzlibrary{arrows,backgrounds}
\tikzset{small/.style={font=\fontsize{9}{9}\selectfont}}
\tikzset{vsmall/.style={font=\fontsize{7}{7}\selectfont}}
\usepackage{graphicx}
\usepackage{wrapfig}
\usepackage{calc}

\def\barr{\begin{array}}
\def\earr{\end{array}}
\def\ben{\begin{equation}}
\def\een{\end{equation}}
\def\bena{\begin{eqnarray}}
\def\eena{\end{eqnarray}}

\def\vol{\rm vol}

\newcommand{\OMs}{{\bf \Omega}} 
\newcommand{\OMp}{K} 

\newcommand{\sperp}{{\scriptscriptstyle\perp}}
\newcommand{\spara}{{\scriptscriptstyle\parallel}}

\newcommand{\starp}{{\ast\!{}_{\scriptscriptstyle\perp}}}
\newcommand{\dperp}{d^{\scriptscriptstyle\perp}}
\newcommand{\ssperp}{{\scriptscriptstyle\perp}}

\newcommand{\blank}{\,\cdot\,}

\newcommand{\spatforms}{\Om_{\!\scriptscriptstyle\perp}} 
\newcommand{\tempforms}{\Om_{\scriptscriptstyle \parallel}} 

\newcommand{\fakeperp}{\fake_{\!\scriptscriptstyle\perp}} 
\newcommand{\fakepara}{\fake_{\!\scriptscriptstyle\parallel}}

\newcommand{\om}{\omega}
\newcommand{\Om}{\Omega}

\newcommand{\R}{{\mathbb R}}

\newcommand{\fake}{\mathcal{T}} 
\newcommand{\ff}{\mathcal{F}} 
\newcommand{\fo}{\mathcal{O}} 

\newcommand{\maps}{\colon}

\def\stackto #1 { \, {\stackrel{#1}{\longrightarrow}}\, }
\def\stackTo #1 { {\stackrel{#1}{\Longrightarrow}} }

\newcommand{\tr}{{\rm tr}}

\newcommand{\Ad}{{\rm Ad}}

\newcommand{\SO}{{\rm SO}}
\newcommand{\so}{\mathfrak{so}}

\newcommand{\ISO}{{\rm ISO}}
\newcommand{\Iso}{\mathfrak{iso}}

\newcommand{\g}{\mathfrak{g}}

\newcommand{\define}[1]{{\bf #1}}

\newcommand{\we}{\wedge}
\renewcommand{\L}{\pounds} 

\newtheorem{thm}{Theorem}

\newenvironment{proof.within.proof}
{\noindent{\it Proof:}}{
\hfill $\Box$ \medskip}

\newcounter{Ccounter} 
\newenvironment{C-list}{  
\begin{list}{{\rm C\arabic{Ccounter}}.}{\usecounter{Ccounter}}
}{\end{list}}

\newcounter{Cpcounter} 
\newenvironment{C'-list}{  
\begin{list}{{\rm C\arabic{Cpcounter}${}'$}.}{\usecounter{Cpcounter}}
}{\end{list}}

\newcommand{\grqc}[1]{\href{http://arxiv.org/abs/gr-qc/#1}{arXiv:gr-qc/#1}}

\newcommand{\arxiv}[1]{\href{http://arxiv.org/abs/#1/}{arXiv:#1}}

\newcommand{\webpage}[1]{{\color{blue}}\url{#1}{\color{blue}}}

\begin{document}

\date{October 3, 2013}

\maketitle

\thispagestyle{empty}

\begin{abstract}
We study the dynamics of gauge theory and general relativity using fields of local observers, thus maintaining local Lorentz symmetry despite a space/time splitting of fields.  We start with Yang--Mills theory, where observer fields are defined as normalized future-timelike vector fields.  We then define observers without a fixed geometry, and find these play two related roles in general relativity:  splitting fields into spatial and temporal parts, and `breaking' gauge symmetry, effectively reducing the spacetime $\SO(n,1)$ connection to an observer-dependent spatial $\SO(n)$ connection.  In both gauge theory and gravity, the observer field reduces the action to canonical form, without using gauge fixing.  In the 4d gravity case, the result is a manifestly  Lorentz covariant counterpart of the Ashtekar--Barbero formulation.  We also explain how this leads geometrically to a picture of general relativity in terms of `observer space' rather than spacetime---a setting where both spacetime symmetry and the dynamical description are simultaneously available.  
\end{abstract}


\section{Introduction: geometrodynamics and observers} 

Geometrodynamics is the picture of general relativity as {\em evolving spatial geometry}.   Since Einstein's equations are not directly about space but about spacetime, the geometrodynamic viewpoint traditionally involves picking an arbitrary `time' coordinate $t$ on spacetime and decomposing equations into a part that describes the field configurations at fixed $t$ and a part describing time evolution.  Two well-studied approaches of this kind are the ADM \cite{adm} and Ashtekar--Barbero \cite{barbero} formulations, which focus respectively on the dynamics of a spatial metric and the dynamics of a spatial $\SO(3)$ connection.  

Unfortunately, the significant advantages of geometrodynamics, including the Hamiltonian description of gravity, usually come at the expense of manifest local Lorentz symmetry.  When local reference frames are bound to the time slicing, and fields put in `time gauge', local Lorentz transformation of frames no longer act in any obvious way.   

This is further complicated in the Ashtekar--Barbero picture, where it is not obvious how local $\SO(3,1)$ transformations should act on the Lie algebra part of the $\SO(3)$ connection.  While Barbero's Hamiltonian formulation derives from a Lorentz-covariant action \cite{holst}, the usual derivation explicitly breaks Lorentz symmetry using a time gauge.  This has resulted in significant controversy, especially in loop quantum gravity, which is founded on the non-covariant Ashtekar--Barbero approach.  Restoring Lorentz covariance in the quantum theory has been a complicated and sensitive issue (see e.g.\ \cite{alexliv,cianmont,geil,rovellispeziale}), suggesting the need for a simple geometric way of maintaining Lorentz covariance from the outset.

Here we present a recent alternative picture of geometrodynamics, relying not on a global time coordinate, but on a {\em local field of observers}.   This  field plays two complementary roles: it splits spacetime into spatial and temporal directions, but also gives an observer-dependent breaking of $\SO(3,1)$ symmetry to $\SO(3)$.  Despite this `broken' symmetry, full Lorentz gauge symmetry is still manifest, since local Lorentz transformations act not only on the physical fields, but also on the observer fields.    

In general relativity and related geometric gauge theories, the appearance of symmetry breaking often hints at geometrical foundations based on {\em Cartan geometry} \cite{broken}.   We show that the symmetry breaking introduced by the observer field gives a `spatial Cartan geometry' on spacetime, which nonetheless respects spacetime Lorentz symmetry.  When the observer field is normal to a hypersurface, this induces a Cartan geometry in the usual sense on the hypersurface, leading to a picture of gravity as evolving Cartan geometries on space, or `{\em Cartan geometrodynamics}'.

The full geometric picture involves symmetry breaking at two levels: first reducing Poincar\'e symmetry to Lorentz symmetry to give spacetime Cartan geometry \cite{symm-mm-cartan}, and then further from Lorentz symmetry to rotational symmetry \cite{lorentz} giving Cartan geometrodynamics.  We will that these geometric pieces fit nicely together in the Cartan geometry of {\em observer space} \cite{lifting}, the space of all possible observers in spacetime.  Observer space provides an setting for general relativity with the advantages of both covariant and canonical approaches.

\section{Observer fields in Yang--Mills theory}
\label{sec:YM}

To help set the stage for general relativity, and build up most of the tools we will need, we first consider a context in which the spacetime geometry is a fixed background structure. 

\subsection{Observers}
\label{sec:observers}

Let $M$ be an ($n$+1)-dimensional manifold with Lorentzian metric $g$, equipped with an orientation and a time orientation.  An \define{observer} in $M$ is a unit future-directed timelike tangent vector, and the space of all observers, the \define{observer space} of $M$, is the unit future tangent bundle $O\subset TM$ \cite{lifting}.  An \define{observer field} $u$ is just a section of $O$---a unit future-timelike vector field.

An observer field $u$ has a corresponding \define{co-observer field} $\hat u:=-g(u,\blank)$, a 1-form whose kernel defines `spatial directions' at each point in $M$.  An $n$-dimensional submanifold of $M$ with tangent spaces in $\ker \hat u$ will be called a \define{spatial hypersurface}; these exist only where the Frobenius condition
\ben
\label{frobenius}
    \hat u \wedge d\hat u=0
\een
holds, 
in which case we can locally write $\hat u = N\, dt$ for some functions $t$ and $N$ corresponding to the time coordinate and lapse function in foliation-based approaches.  \cite{lorentz}

Whether or not an observer field induces a spatial foliation, it gives a canonical way to split any physical field on $M$ into `temporal' and `spatial' parts.  In particular, for the complex of differential forms $\Om^\bullet(M)$, we define the \define{spatial forms} $\spatforms^\bullet$ to be the kernel of interior multiplication $\iota_u$, and \define{temporal forms} $\tempforms^\bullet$ to be the kernel of wedging with $\hat u$.  This puts an additional grading on $\Om^\bullet(M)$:
\[
  \Om^\bullet (M) = \spatforms^\bullet(M) \oplus \tempforms^\bullet(M),
\]
where $\spatforms$ has grade 0 and $\tempforms$ has grade 1. 
The corresponding projections into spatial and temporal parts are  
\ben
X^{\perp}:=\iota_{u}(\hat{u}\wedge X)\,, \quad 
X^{\parallel}:=\hat{u} \we \iota_{u} X. 
\een
The spatial projection is a homomorphism; the temporal projection is a derivation. 

Likewise, observers have preferred spatial and temporal derivatives.  The natural time derivative for the observer field $u$ is the Lie derivative $\L_u =  \iota_u d + d \iota_u$, and the Lie derivative of a spatial form is spatial, since $\L_u$ commutes with $\iota_u$.  The \define{spatial differential}
\ben
 \dperp X = dX -  \hat{u}\wedge \L_u X 
\een
is always a graded derivation, but squares to zero precisely when the Frobenius condition (\ref{frobenius}) holds; in this case it is indeed the usual spatial differential on each leaf of the corresponding foliation. 

We also define a \define{spatial Hodge star} $\starp\maps \spatforms^p \to \spatforms^{n-p}$, related to the usual spacetime Hodge star operator $\ast\maps \Omega^p \to \Omega^{n+1-p}$ by
\[
   \starp \! X := - {\ast}(\hat u \wedge X) \quad \text{and} \quad 
  \hat u \wedge {\starp X} = (-1)^p {\ast} X \quad \text{for } X \in \Omega^p_\perp.
\]
This works because the spacetime Hodge star automatically interchanges spatial and temporal forms for any observer.  Just as $X\we \ast Y = \langle X,Y\rangle \vol$, where $\vol$ is the volume form on $M$, we have $X\we \starp Y = \langle X,Y\rangle \iota_u\vol$ whenever $X$ and $Y$ are both spatial. 

\subsection{Yang--Mills}

We can apply all of this, for example, to Yang--Mills theory.  Introducing an observer field, the action $S[A]= - \int \tr(F\we {\ast F})$ now falls almost effortlessly into a familiar-looking form: 
\ben
\label{YMaction}
  S[A] = \int \hat u \we \tr \big( E\we \starp\! E -  B\we \starp\! B \big).   
\een 
Here $E:=-\iota_u F$ and $B:=F^{\ssperp}$ are the nonabelian electric and magnetic fields, so that $F= B-\hat u \wedge E$.  For Minkowski spacetime with $u=\partial_t$ and $\hat u = dt$, we thus get the usual action $\int dt (E^2 - B^2)$. 
 
Further analysis of Yang--Mills theory using observer fields yields familiar formulas, but with some unexpected new terms involving derivatives of $\hat u$.  For example, writing $A = a - \hat u \phi$, where $a:= A^\perp$ is the nonabelian magnetic vector potential and $\phi:= -\iota_u A$ is the nonabelian electric potential, we find
\[
   B = d^{\scriptscriptstyle \perp\!} a+a\we a - (\dperp \hat u) \phi,
   \quad 
   E= -\dperp_a \phi  - \L_u a + (\L_u \hat u)\phi 
\]
which are familiar except for the $\dperp \hat u$ and $\L_u \hat u$ terms.  In fact, $\dperp \hat u$ vanishes whenever the Frobenius integrability condition holds, and $\L_u \hat u$ vanishes whenever $u$ is a Killing vector field.  

Writing $E$ and $B$ in terms of their potentials, action again takes familiar form, this time revealing the Hamiltonian analysis:
\[
 S= - \int \hat u \we \tr \big( E\we \starp\! \L_u a +\cdots \big)
\]
We have written just the first term, from which we can see that $E$ is the momentum of the vector potential $a$, as usual.  The omitted terms give the usual constraints, though again with new terms proportional to derivatives of $\hat u$. 

The observer-based decomposition of the Yang--Mills equations $d_A F = 0$ and $d_A {\ast}F=0$ is also straightforward:
\[
\begin{array}{lllll}
\text{\tiny\sf Maxwell-like} & \text{\tiny\sf nonabelian self-sourcing}  &  {\tiny \txt{\sf terms that \\ \sf vanish if Frobenious\\ \sf condition (\ref{frobenius}) holds} }&  {\tiny \txt{\sf terms that \\ \sf vanish if $u$ is a \\ \sf Killing vector field}}\\[1em]
 \dperp B &+ [a, B] &- \dperp \hat u \we E  & &= 0 \\
 \dperp E + \L_u B &+[a,E] - [\phi,B] &&- \L_u \hat u \we E  &= 0\\
 \dperp \starp E &+ [a,\starp E] &+ \dperp \hat u \we \starp B   & &= 0 \\
 \dperp \starp B - \L_u \starp E &+ [a,\starp B] + [\phi,\starp E] &&- \L_u \hat u \we \starp B  &= 0 
\end{array}
\]
The real reason for the terms involving $\dperp \hat u$ and $\L_u \hat u$ is that the observer-based decomposition has local Lorentz symmetry much less evident in the foliation-based approach. 

\subsection{Local Lorentz symmetry}

If spacetime is Lorentzian, then physics should be invariant under local Lorentz transformations relating local observers to each other.  An \define{observer transformation} \cite{lorentz} is simply a gauge transformation $\lambda\maps O\to O$, acting on sections by composition: $u\mapsto \lambda\circ u$.  In a local trivialization, this amounts to left multiplication by an $\SO(n,1)$-valued function.  
Any two observer fields are related by some observer transformation, and obviously all of the equations we have written are invariant under observer transformations. 

Foliation invariance is a special case of invariance under observer transformations.  An observer field satisfying the Frobenius condition (\ref{frobenius}) is equivalent to a spacelike foliation of $M$.   Thus, any two spacelike foliations are related by some observer transformation. 
However, there is no subgroup of the group of gauge transformations that preserves the property (\ref{frobenius}) of being a foliation, since this depends on the initial section $u$.  Thus the transformations going from one {\em foliation} to another form  not a group but a groupoid---a sub-groupoid of the transformation groupoid.

\section{Observers in general relativity}

Two issues arise when we attempt a similar decomposition of general relativity.  First, we face a dilemma of priority:  we want the dynamics of gravity as seen by an observer, but the very definition of an observer requires the metric, which is determined by the dynamics. Second, unlike in Yang--Mills theory, where the gauge group is unaffected by the decomposition, here we start with an $\SO_o(n,1)$ connection on spacetime, but each observer should see a spatial $\SO(n)$ connection.  These issues are related, and we deal with each in turn.  

\subsection{Observers without a background geometry} 

With no fixed geometry on $M$, an ($n$+1)-dimensional manifold, we will create a `fake' geometry for observers to live in.  Pick a vector bundle $\fake$ over $M$, isomorphic to the tangent bundle $TM$, but so far not in any specified way.  We call $\fake$ the \define{fake tangent bundle} and equip it with all of the structure we demanded of $TM$ in Sec.~\ref{sec:YM}, namely a fixed `metric' (a smoothly varying Lorentzian inner product on fibers) as well as an orientation and time orientation.  

A \define{fake observer} on $M$ is a unit future-directed timelike element of $\fake$, and the space of all fake observers is the \define{fake observer space} $\fo\subset \fake$ \cite{lifting}.  A \define{fake observer field} $y$ is just a section of $\fo$. 
Of course, $y$ lets us do the same things with $\fake$ that $u$ lets us do with $TM$, and $\hat u$  with $T^\ast\!M$. 

In particular, $y$ splits  $\fake$ into a part $\fakeperp$ orthogonal to $y$ and a part $\fake_\spara$ parallel to $y$.  More generally, it splits the exterior bundle $\Lambda^\bullet \fake$ into a fiberwise direct sum
\[
   \Lambda^\bullet \fake = \Lambda_\sperp^\bullet \fake \oplus_M \Lambda_\spara^\bullet\fake.
\]
Here  
$
\Lambda_\sperp^\bullet \fake = \Lambda^\bullet\fakeperp 
$
is generated by wedge products of vectors orthogonal to $y$, while $\Lambda_\spara^\bullet\fake$ is the kernel of the exterior product with $y$.    Since fibers of $\fake$ are oriented  Lorentzian inner product spaces, there is a Hodge star on $\Lambda^\bullet \fake$, which we write as `$\star$' to distinguish it from the Hodge star `$\ast$' on differential forms.  
Just as $\ast$ turns spatial forms into temporal forms,     
we also have 
\ben
{\star}(\Lambda^{p}_\sperp\fake) = \Lambda_\spara^{n+1-p}\fake.
\label{starLambda-fake}
\een

Fake observers become `real' once we introduce a \define{coframe field}---a vector bundle isomorphism 
\[
   e\maps TM \stackrel\sim\to \fake.
\]
A coframe field in general relativity determines the spacetime metric by  pullback, but here it also conspires with the fake observer field $y$ to give an observer field $$u= e^\ast y.$$ 
This observer field splits fields into spatial and temporal parts, just as in Sec.~\ref{sec:observers}. 

Most notably, the coframe field,  itself a $\fake$-valued 1-form, splits into spatial and temporal parts:
\ben
   e = E + \hat u y.
\label{spatial-coframe}
\een
A crucial observation is that no further splitting using $\fake=\fakeperp \oplus \fakepara$ is needed: the spatial 1-form  $E$ already takes values in $\fakeperp\subset\fake$, since $y$ by definition, takes values in $\fakepara$.   On any spatial hypersurface $S$, $E$ restricts to an isomorphism $E\maps TS \to \fakeperp|_S$, so $E$ is justly called the \define{spatial coframe field}.

\subsection{Breaking Lorentz symmetry} 

Since the fake observer field $y$ determines the spatial sub-bundle $\fakeperp \subset \fake$, it also `breaks' local Lorentz symmetry to rotational symmetry.   It is easiest to see how all physical fields are affected if we write this symmetry breaking in the language of principal bundles.

The \define{fake frame bundle} $\ff$ is the principal $\SO_o(n,1)$ bundle of orthonormal frames in $\fake$ respecting the orientation and time orientation, where $\SO_o(n,1)$ is the connected Lorentz group.  A fake observer field $y$ is a section of the fake observer space $\fo$, but this just the associated bundle $\fo\cong \ff\times_{\SO_o(n,1)} H^n$, where $H^n \cong \SO_o(n,1)/\SO(n)$ is hyperbolic space.  Such a section is the same as a reduction of $\ff$ to a principal $\SO(n)$ bundle $\ff_y\to M$, given by pullback:
\[
\begin{tikzpicture}[xscale=1.5,yscale=1.5,>=stealth']
  \node (O) at (1,0) {$\fo$};
  \node (M) at (0,0) {$M$}  
    edge [->] node [above,small] {$y$} (O);
  \node (F) at (1,1) {$\ff$}
    edge [->] (O);
  \node (Fy) at (0,1) {$\ff_y$}
    edge [->] (F)
    edge [->] (M);
\end{tikzpicture}
\]

 Thanks to this reduction, any physical field living in a representation of $\SO_o(n,1)$ now splits into representations of $\SO(n)$.  In particular, the fundamental and adjoint representations each split into two irreducible representations of $\SO(n)$:
\ben
\begin{array}{ccccl}
\R^{n,1} & \cong & \R^n &\oplus& \R \\
\so(n,1)&\cong& \so(n) &\oplus& \R^n
\end{array}
\label{rep-split}
\een
The two representations labeled $\R^n$ are canonically isomorphic.  The group $\SO(n)$ is the stabilizer of a unique `observer' $\xi\in \R^{n,1}$, a unit timelike vector with positive `time' component.  The $\R^n$ in the first line of (\ref{rep-split}) is the orthogonal complement of $\xi$; the map $\R^n\to \so(n,1)$ given by $ v \mapsto \eta(\xi,\blank)v - \eta(v,\blank)\xi$ is an intertwiner of $\SO(n)$ representations and its image is the $\R^n$ in the second line of (\ref{rep-split}).

\subsection{Lorentz equivariant spatial connections}

We can now decompose a connection $\omega$ on $\ff$ in two ways, both into spatial and temporal parts and via the reduction to $\SO(n)$ symmetry.  Doing the spatiotemporal splitting first, write
\[
  \omega=\Omega+\hat{u} \Xi
\]
where $\Omega = \om^\sperp$ and $\Xi = \iota_u \omega$.  
Then, since $\Om$ can be thought of as an $\so(n,1)$-valued 1-form on $\ff$, we can pull it back to $\ff_y$, where it splits into two irreducible $\SO(n)$ subrepresentations of $\so(n,1)$:
\ben
\label{spatial-connection-split}
\begin{array}{ccccc}
 \Om &=& \OMs &+& \OMp\\
 && \text{\scriptsize $\so(n)$ part} && \text{\scriptsize $\R^n$ part} 
\end{array}
\een
corresponding to infinitesimal rotations and boosts, from the perspective of the observer.  We call $\OMs$ the \define{\boldmath spatial $\SO(n)$ connection}.  It is a spatial form, $\iota_u \OMs=0$, and remains so under local Lorentz transformations acting both on $\OMs$ and $u$.  It restricts to a connection on any spatial hypersurface.  

By splitting the torsion $d_\om e$ into spatial and temporal parts, and further into representations of $\SO(n)$, one can show that $\dperp_\OMs E$ vanishes whenever $d_\om e$ does.  In particular, when the Frobenius condition holds, $\OMs$ restricts to the unique torsion free connection on any spatial hypersurface.

\subsection{Palatini} 

Using all of this, let us now introduce a fake observer field in the Palatini action for general relativity.  The action in $n+1$ dimensions can be written, up to a constant factor, as
\ben
\label{palatini}
    S[e,\omega] = \int     \tr \Big({\star}({\color{gray}\underbrace{{\color{black}e\wedge\cdots \wedge e}}_{n-1}}) \wedge R 
       \Big), 
\een
which resembles Yang--Mills theory, since both ${\star}(e\we \cdots \we e)$ and $R$ live in $\so(n,1)$.  More precisely, ${\star}(e\we \cdots \we e)$ is a $\Lambda^2\fake$-valued $(n-1)$-form, but $\Lambda^2\fake\cong\Ad(\ff)$, thanks to the isomorphism $\Lambda^2\R^{n,1}\cong\so(n,1)$ between bivectors and infinitesimal pseudorotations.  Thus both forms take values in $\Ad(\ff)$ and the trace is exactly as in Yang--Mills theory.  

Obviously nothing changes about the theory if we introduce a fake observer field $y$, a section of $\ff\times_{\SO_o(n,1)} H^n$, but leave the formula for the action unchanged, defining
\[
  S[e,\om,y] := S[e,\om]. 
\]
Using the observer field $u=e^\ast y$, we can split fields into spatial and temporal parts.  This breaks the action into a number of terms, of which we write just the first:
\ben
    \tr \Big({\star}({\color{gray}\underbrace{{\color{black}e\wedge\cdots \wedge e}}_{n-1}}) \wedge R 
       \Big)  =   \hat u \we \tr \Big({\star}({\color{gray}\underbrace{{\color{black} E \wedge\cdots \wedge E}}_{n-1}}) \wedge \L_u \Omega
     +\cdots   \Big). 
\label{palatini-observer1}
\een
The remaining terms are straightforward to work out, but unimportant for present purposes.  

This simplifies when we also split fields according to representations of $\SO(n)$,  because $\L_u\Om$ lives in $\so(3,1) = \so(3) \oplus \R^n$, whereas $\star(E\we \cdots \we E)$ lives just in $\R^n$.  More precisely, since $E$ is $\fakeperp$-valued, $(E\we \cdots \we E)$  is $\Lambda^{n-1}\fakeperp$-valued and, by (\ref{starLambda-fake}), ${\star} (E\we \cdots \we E)$ is $\Lambda_\spara^{2}\fake$-valued.  But under the isomorphism $\Lambda^2\fake\cong \ff_y \times_{\SO(n)}\so(n,1)$, $\Lambda_\spara^{2}\fake$ is identified with $\ff_y\times_{\SO(n)} \R^n$.  Thus, in the first term of the action, we can replace $\Omega$ with its $\R^n$ part: 
\ben
    S[e,\omega] =  \int     \hat u \we \tr \Big({\star}({\color{gray}\underbrace{{\color{black} E \wedge\cdots \wedge E}}_{n-1}}) \wedge \L_u K
     +\cdots   \Big). 
\label{palatini-observer2}
\een

We have arrived at a hybrid of canonical and covariant formulations.   Simply introducing the field $y$, the action falls into the form $\int dt\;(p\dot q + \cdots)$, from which we can read off the canonical momenta and the constraints.  In particular the $\so(n)$ part $\OMs$ of the spatial connection (\ref{spatial-connection-split}) is nondynamical, and only the $\R^n$ part $K$ has nonvanishing momentum.  This agrees with standard Hamiltonian analysis of the Palatini action (see e.g.\ \cite[Sec.~4.2]{thiemann}) where $K$ is identified as the extrinsic curvature of a spatial hypersurface.   On the other hand, since Lorentz transformations act on all fields including $y$, we maintain Lorentz covariance.

\subsection{Ashtekar--Barbero} 
In 3+1 dimensions, a modification \cite{holst} of the action (\ref{palatini}) is possible, since  $\star$ maps $\Lambda^2\fake \cong \Ad(\ff)$ 
to itself:
\ben
\label{holst-action}
    S[e,\omega] = \int     \tr \Big(({\star}+{\textstyle\frac 1\gamma})(e\wedge e) \wedge R 
       \Big).
\een
Splitting the fields as before into spatial and temporal parts and by $\SO(3)$ representations, the analog of (\ref{palatini-observer2}) is
\ben
\label{holst-observer}
S  =  \int \hat{u}\wedge\tr\Big({\star}(E\wedge E)\wedge \L_u(K - {\textstyle\frac 1\gamma}{\star}\OMs) + 
\cdots
\Big)
\een
Whereas in (\ref{palatini-observer2}) ${\star}(E\we E)$ was the momentum of $K$, it is now the momentum of $-{\textstyle\frac 1\gamma}{\star}\sf A$, where 
\ben
         {\sf A}=  \OMs + \gamma {\star \OMp} .
\label{ABconn}
\een
is a spatial $\SO(3)$ connection, the analog in our framework of the \define{Ashtekar--Barbero connection}.  It is a Lorentz-covariant spatial connection in the same sense as $\OMs$, but differs from $\OMs$ in its torsion and curvature.  In particular, since $d_{\sf A} = d_\OMs - \gamma[{\star}K,\blank]$, the torsion $d_{\sf A} E$ is propotional to ${\star}K$, which in the case of a foliation is just the extrinsic curvature of the observer field (cf.~\cite{montesinos}).  

The Hamiltonian analysis following from (\ref{holst-observer}) is similar to the standard analysis using gauge fixing \cite{holst}, but maintaining Lorentz symmetry using observer fields also avoids the second class constraints that are inevitable in gauge-fixed approaches.  Details can be found in our previous work \cite{lorentz}.  Here, we continue instead with the geometric interpretation.  

\section{Cartan geometrodynamics}

\subsection{Evolving spatial Cartan geometries}
Cartan geometry is ubiquitous in gauge-theoretic formulations of gravity \cite{symm-mm-cartan}.  Its appearance in spacetime physics is signaled by a connection with values in some group $G$, but with gauge symmetry only under a closed subgroup $H$, and where the homogeneous space $G/H$ has the same type of geometry as spacetime, usually a Lorentzian manifold of the same dimension.  In general, a \define{Cartan geometry} modeled on $G/H$ is a principal $H$ bundle $Q\to M$ equipped with an $H$-equivariant $\g$-valued 1-form that is a linear isomorphism at each point of $Q$ and restricts to the Maurer--Cartan form on vertical vectors.

A key example in gravitational physics comes from treating the connection $\omega$ and coframe field $e$ in (\ref{palatini}) as the $\so(n,1)$ and $\R^{n,1}$ parts of a unified $\Iso(n,1)$-valued connection.  This connection can be viewed as 1-form on the principal $\SO(n,1)$ bundle $\ff$, and this gives a Cartan geometry modeled on Minkowski spacetime $\ISO(n,1)/\SO(n,1)$.

The `spatial geometry' studied here nicely parallels this example.  Just as $\omega + e$ gives a Cartan geometry modeled on Minkowski spacetime, $\OMs + E$ gives a `spatial' Cartan geometry modeled on Euclidean space.  More precisely:
\begin{thm}
Let $(\ff\to M,\omega+e)$ be a Cartan geometry with model $\ISO_o(n,1)/\SO_o(n,1)$; let $y$ be a fake observer field with corresponding observer field $u=e^\ast y$.     
If $S$ is any integral hypersurface of $\ker \hat u$ 
then $(\ff_y|_S\to S, \OMs + E)$ is a Cartan geometry with model $\ISO(n)/\SO(n)$, where $\OMs$ is the spatial $\SO(n)$ connection defined in {\rm (\ref{spatial-connection-split})} and $E$ is the spatial coframe {\rm (\ref{spatial-coframe})}.  In the case $n$=3, the same is also true if we replace $\OMs$ by the Ashtekar--Barbero connection $\sf A$ {\rm (\ref{ABconn})}.
\end{thm}

In particular, if $S=S_t$ is the level set of a time function $t$ on spacetime, then each spatial slice $S_t$ becomes a spatial geometry.  There is a canonical correspondence between metrics and torsion-free Cartan geometries modeled on Euclidean space.

\subsection{Symmetry breaking and observer space}

It is often useful to think about Cartan geometry in physics in terms of spontaneous symmetry breaking.  The reason is that a Cartan connection with model $G/H$ can be viewed as an ordinary Ehresmann connection on a principal $G$ bundle, pulled back along a reduction to an $H$ bundle.

In our case, although we have used the observer field to break symmetry, the relation to spatial Cartan geometry is not immediate.   In particular, the theorem in the previous section refers to spatial Cartan geometry modeled on {Euclidean} space $\ISO(n)/\SO(n)$, but $\ff_y$ is instead a reduction of the $\SO(n,1)$ bundle $\ff$; we have no obvious $\ISO(n)$ bundle in sight.   Of course, we can {extend} $\ff_y$ to an $\ISO(n)$ bundle, and extend the Cartan connection $\OMs+E$ to an Ehresmann connection on the extension, but this seems less than natural.   It would be nice to see the spatial Cartan geometry coming directly from $\ISO(n)$ symmetry breaking.  

It helps to think more carefully about observers' role in symmetry breaking.  
In Minkowski spacetime, there are two equivalent ways to specify an observer.  One can first pick a point in spacetime, reducing $\ISO(n,1)$ symmetry to $\SO(n,1)$, and then a unit timelike vector there, reducing further to $\SO(n)$.  Equivalently, one can pick first a spacelike hyperplane, or surface of simultaneity, reducing symmetry to $\ISO(n)$, and then pick the location of the observer within that hyperplane, reducing further to $\SO(n)$.    The features and their corresponding stabilizers are:
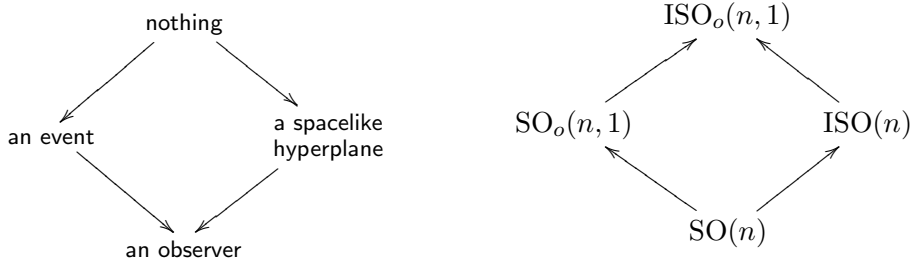
\[
\xymatrix@C=.2cm{
&{\footnotesize \txt{\sf nothing}} \ar[dl]\ar[dr] & \\
{\footnotesize \txt{\sf an event}}\ar[dr] &&{\footnotesize \txt{\sf a spacelike \\ \sf hyperplane}}\ar[dl] \\
& {\footnotesize \txt{\sf an observer}} 
}
\qquad\qquad 
\xymatrix@C=.2cm{
&\ISO_o(n,1)& \\
\SO_o(n,1)\ar[ur] && \ISO(n)\ar[ul] \\
& \SO(n) \ar[ul]\ar[ur]
}
\]
The end result is the same, but the intermediate steps in the two branches are quite different.  

Moreover,  breaking symmetry all the way from $\ISO_o(n,1)$ down to $\SO(n)$, the Poincar\'e Lie algebra $\Iso(n,1)$ splits into four irreducible $\SO(n)$ representations.  This can be viewed either in terms of spacetime or the space of spacelike hyperplanes, and the relationship between these two perspectives is best explained with a diagram:   
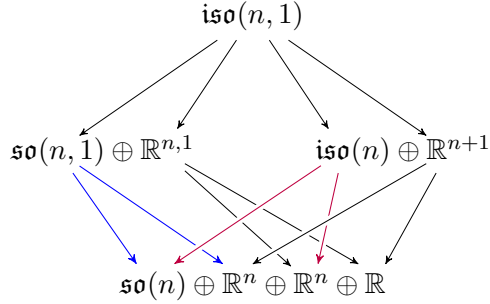
\[
\begin{tikzpicture}[xscale=2,yscale=1.8,>=stealth']
  \node (h) at (-1,0) {$\so(n,1) \oplus \R^{n,1}$};
  \node (h') at (1,0) {$\Iso(n) \oplus \R^{n+1}$};
  \node (k) at (0,-1) {$\so(n) \oplus \R^n \oplus \R^n \oplus \R$};
  \begin{scope}
  \node [minimum size = 1.5em] (k1) at (-.67,-1) {};
  \node [minimum height = 1.5em, minimum width = 1em] (k2) at (-.1,-1) {};
  \node [minimum size = 1.5em] (k3) at (.4,-1) {};
  \node [minimum height = 1.5em, minimum width = 1em] (k4) at (.8,-1) {};
  \node [minimum size = 1.5em] (h1) at (-1.3,0) {}
     edge [->,color=blue] (k1) 
     edge [->,color=blue] (k2.north west);
  \node [minimum size = 1.5em] (h2) at (-.6,0) {}
     edge [->,color=black,very thin] (k3) 
     edge [->,color=black,very thin] (k4.north west);
  \node [minimum size = 1.5em] (h'1) at (.6,0) {}
     edge [white,line width=1mm] (k1.north east) edge [->,color=purple] (k1.north east)
     edge [white,line width=1mm] (k3) edge [->,color=purple] (k3);
  \node [minimum size = 1.5em] (h'2) at (1.3,0) {}
     edge [white,line width=1mm] (k2.north east) edge [->,color=black,very thin] (k2.north east)
     edge [->,color=black,very thin] (k4);
  \end{scope};
  \node (g) at (0,1) {$\Iso(n,1)$}
     edge [->,color=black,very thin] (h1) 
     edge [->,color=black,very thin] (h2)
     edge [->,color=black,very thin] (h'1) 
     edge [->,color=black,very thin] (h'2);
\end{tikzpicture}
\]
The adjacent pairs of arrows are the two projections of a binary direct sum of representations; in the bottom row, each direct summand is the intersection of the two representations pointing to it.

Let us write the decomposition of just the spatial part of the spacetime Cartan connection:
\[
\begin{array}{r@{\;}c@{\;}c@{\;}c@{\;}c@{\;}c@{\;}c@{\;}c@{\;}c@{\;}c@{\;}c}
& & \so(n) &\oplus &\R^n &\oplus &\R^n &\oplus &\R  \\
(\Om+e)^\perp &=&\OMs & + & K & + & E & +& 0 & \phantom{(\Om+e)^\perp} & \phantom{=}
\end{array}
\]
From the spacetime perspective, it seems rather ad hoc to join $\OMs$, the spatial $\so(n)$ part of a Lorentz connection, with $E$, the spatial part of a coframe, into an $\ISO(n)$ connection.  But on the other side of the picture, based on the space of spatial hyperplanes, these pieces seem to come naturally from broken $\ISO(n)$ symmetry.  In particular, Cartan geometrodynamics is clearer when we consider both levels of symmetry breaking at once.

What is the {\em geometric} meaning of breaking symmetry all the way from $\ISO_o(n,1)$ to $\SO(n)$? Evidently, this should lead to Cartan geometry modeled on the homogeneous space
\[
    O(\R^{n,1}) \cong \ISO_o(n,1)/\SO(n)
\]
which is the (2$n$+1)-dimensional {\em observer space} of Minkowski spacetime.  We call a Cartan geometry modeled on $O(\R^{n,1})$ an {\bf observer space geometry} \cite{lifting}. While observer space has played a mostly superficial role in this paper, as the bundle whose sections are observer fields, we now see that the full geometric picture of covariant canonical gravity gives observer space a life of its own.  

In fact, the spacetime Cartan geometry $(\ff,\om+e)$ on $M$ induces an observer space geometry on  $\fo$.  Different descriptions of the same physical situation in terms of Cartan geometrodynamics are given by pulling this Cartan geometry back along sections of the bundle $\fo \to M$.  In fact, we almost saw this this before, when we noted how $y$ breaks Lorentz symmetry: 
\[
\begin{tikzpicture}[xscale=1.5,yscale=1.5,>=stealth']
  \node (O) at (1,0) {$\fo$};
  \node (M) at (0,0) {$M$}  
    edge [->] node [above,small] {$y$} (O);
  \node (F) at (1,1) {$\ff$}
    edge [->] (O);
  \node (Fy) at (0,1) {$\ff_y$}
    edge [->] (F)
    edge [->] (M);
\end{tikzpicture}
\]
The bundle on the right is the principal $\SO(n)$ bundle over observer space, on which the canonical observer space geometry is built.   Each `canonical' description of a solution of general relativity corresponds to one such pullback square.  However, it is the observer space geometry itself that is truly `canonical' in the mathematical sense of being constructed without arbitrary choices \cite{essay,lifting}.

\section*{Acknowledgments}
This paper is based largely on work with Steffen Gielen \cite{lorentz,essay,lifting}, to whom I am also thankful for comments and suggestions.


\begin{thebibliography}{99}

\small
\itemsep 0em

\bibitem{alexliv} S. Alexandrov and E. R. Livine, SU(2) loop quantum gravity seen from covariant theory, {\sl Phys. Rev. D\ }{\bf 67}, 044009 (2003). \grqc{0209105}



\bibitem{adm} R. Arnowitt, S. Deser, and C. W. Misner, The dynamics of general relativity, in {\it Gravitation: an introduction to current research}, edited by L. Witten (Wiley, New York, 1962). Reprint available as \grqc{0405109}.

\bibitem{barbero} J.\ F.\  Barbero G., 
Real Ashtekar variables for Lorentzian signature space times, 
{\sl Phys. Rev. D\ }{\bf 51}, 5507-5510 (1995). \grqc{9410014}.

\bibitem{cianmont} F. Cianfrani and G. Montani, Towards loop quantum gravity without the time gauge, {\sl Phys. Rev. Lett.\ }{\bf 102}, 091301 (2009). \arxiv{0811.1916}

\bibitem{geil} M. Geiller, M. Lachi\`{e}ze-Rey, K. Noui, and F. Sardelli, A Lorentz-covariant connection for canonical gravity, {\sl SIGMA\ }{\bf 7}, 083 (2011). \arxiv{1103.4057}.

\bibitem{lorentz} S.\ Gielen and D.\ K.\ Wise, Spontaneously broken Lorentz symmetry for Hamiltonian gravity, {\sl Phys.\ Rev.\ D} {\bf 85} (2012) 104013, \arxiv{1111.7195}.

\bibitem{essay} S.\ Gielen and D.\ K.\ Wise, Linking covariant and canonical general relativity via local observers, {\sl Gen.\ Relativ.\ Gravit.\ }{\bf 44} (2012) 3103-3109, \arxiv{1206.0658}.

\bibitem{lifting} S.\ Gielen and D.\ K.\ Wise, Lifting general relativity to observer space, \href{http://link.aip.org/link/doi/10.1063/1.4802878}{{\sl J.\ Math.\ Phys.\ }{\bf 54} (2013) 052501}, \arxiv{1210.0019}.

\bibitem{holst} S. Holst, Barbero's Hamiltonian derived from a generalized Hilbert-Palatini action, {\sl Phys. Rev. D\ }{\bf 53}, 5966 (1996). \grqc{9511026}.


\bibitem{montesinos} M.\ Montesinos, Connection with torsion from Ashtekar--Barbero connection, {\sl 
J.\ Math.\ Phys.\ }{\bf 40} 4084 (1999). 


\bibitem{rovellispeziale} C. Rovelli and S. Speziale, Lorentz covariance of loop quantum gravity, {\sl Phys. Rev. D\ }{\bf 83}, 104029 (2011). \arxiv{1012.1739}



\bibitem{thiemann} T. Thiemann, {\it Modern Canonical Quantum General Relativity} (Cambridge University Press, Cambridge, 2008).

\bibitem{symm-mm-cartan} D.\ K.\ Wise, Symmetric space Cartan connections and gravity in three and four dimensions, {\sl SIGMA} {\bf 5} (2009) 080, \arxiv{0904.1738}; 
MacDowell-Mansouri gravity and Cartan geometry, {\sl Class.\ Quant.\ Grav.} {\bf 27} (2010) 155010, \grqc{0611154}.
%
\bibitem{broken} D.\ K.\ Wise, The geometric role of symmetry breaking in gravity, {\sl J.\ Phys.:\ Conf.\ Ser.\ }{\bf 360} (2012) 012017. \arxiv{1112.2390}.


\end{thebibliography}
\end{document}